\documentclass[journal]{IEEEtran}

\usepackage[nolist]{acronym}
\usepackage{amsmath}
\usepackage[table,xcdraw,dvipsnames]{xcolor}

\usepackage{balance}
\usepackage{tabularx}
\usepackage{subcaption}
\usepackage{multirow,array}
\usepackage{url}
\usepackage{textcomp}

\usepackage{easyReview}
\ifCLASSINFOpdf
\usepackage{graphicx}
\fi
\renewenvironment{IEEEbiography}[1] 
{\IEEEbiographynophoto{#1}}
{\endIEEEbiographynophoto}

\begin{document}
\title{Networked Twins and Twins of Networks: an Overview on the Relationship Between Digital Twins and 6G}

\author{\IEEEauthorblockN{Hamed Ahmadi \IEEEmembership{Senior Member,~IEEE}, Avishek Nag \IEEEmembership{Senior Member,~IEEE}, Zaheer Khan \IEEEmembership{Senior Member,~IEEE}, Kamran Sayrafian \IEEEmembership{Senior Member,~IEEE}, Susanto Rahadrja \IEEEmembership{Fellow,~IEEE}
\thanks{H. Ahmadi is with the University of York, UK. A. Nag is with University College Dublin, Ireland. Z. Khan is with 6G centre, University of Oulu, Finland. K. Sayrafian is with National Institute of Standards and  Technology, USA. S. Rahardja is with Institute for Infocomm Reseasech, Singapore.}
}

}

\maketitle

\begin{abstract}
Digital Twin (DT) is a promising technology for the new immersive digital life with a variety of applications in areas such as Industry 4.0, aviation, and healthcare. Proliferation of this technology requires higher data rates, reliability, resilience, and lower latency beyond what is currently offered by 5G. Thus, DT can become a major driver for 6G research and development. Alternatively, 6G network development can benefit from Digital Twin technology and its powerful features such as modularity and remote intelligence. Using DT, a 6G network (or some of its components) will have the opportunity to use Artificial Intelligence more proactively in order to enhance its resilience. DT's application in telecommunications is still in its infancy. In this article we highlight some of the most promising research and development directions for this technology.
\end{abstract}
\section{Introduction}

As commercial deployments of \ac{5G} continues in several countries, 
researchers in the industry and academia have started to focus on the \ac{6G}. 
A range of new technologies such as use of higher frequency bands (THz), Orbital Angular Momentum (OAM) multiplexing, and intelligent surfaces have been proposed for this purpose. In addition, innovative paradigms like integration of terrestrial and satellite networks, massive use of \ac{ML} and \ac{AI}, and quantum and molecular communications for the physical, \ac{MAC}, and network layers are also under development. All of these upcoming technologies and paradigms can be considered as enablers of \ac{6G} 
\cite{viswanathan2020communications,saad2019vision,5g-ppp}. However, researchers are still debating on the importance or potential role of each one of the aforementioned technologies in \ac{6G}. For example, Viswanathan and Morgensen \cite{viswanathan2020communications} believe that \acp{UAV} and cell-free communications belong to the \ac{5G} era, whereas Quantum, Visible Light, and molecular communications are more long-term technologies which will not be mature enough even for \ac{6G} implementation. Since 6G is not fully defined yet, these views are not necessarily shared by other researchers.

\begin{table}[h]\centering
\caption{KPIs of 5G and 6G \cite{viswanathan2020communications,saad2019vision,5g-ppp}.}\label{Table:KPI}
\begin{tabular}{|l|l|l|}
\hline
\textbf{KPIs}                & \textbf{5G}     & \textbf{6G}               \\ \hline
Data rate           & 10+Gbs & 100 Gbs          \\ \hline
Delay               & 1 ms   & 0.5 ms           \\ \hline
Position precision  & meter  & centimeter       \\ \hline
Reliability         & 99.9\% & 99.999\%         \\ \hline
Device intensity    & 1 Million/$\text{Km}^2$ & 10 Million/$\text{Km}^2$         \\ \hline
Spectral efficiency &    -    & 3x more than 5G  \\ \hline
Energy Efficiency   &    -    & 10x more than 5G \\ \hline
\end{tabular}
\end{table}

Unlike the disagreement on the exact technologies that are needed for the development of \ac{6G}, there are more productive discussions and close to agreement on \ac{6G} \acp{KPI}. Table \ref{Table:KPI} shows the targeted \acp{KPI} of \ac{6G} in comparison to \ac{5G} which are gathered from \cite{viswanathan2020communications,saad2019vision,5g-ppp}.

These \acp{KPI} are generally defined to satisfy the requirements of the driving applications of \ac{6G} such as connected robotics, autonomous systems, \ac{AR}/\ac{VR}/\ac{MR}, Blockchain and Trust technologies, and wireless brain-computer interfaces \cite{saad2019vision}. Some of these applications like connected robotics or \ac{AR}/\ac{VR}/\ac{MR} have been considered in \ac{5G} but their massive use could demand higher levels of KPIs beyond what is achievable by \ac{5G} \cite{saad2019vision}. For example, applications like autonomous driving and immersive \ac{AR}/\ac{VR}/\ac{MR} with high definition $360^{\circ}$ video streaming for navigation and/or entertainment are expected to require $99.999\%$ reliability and one millisecond latency \cite{park2019wireless}.

The technologies and driving applications of \ac{6G} enable an environment where a comprehensive digital representation of the physical world can be created and maintained through \acp{DT} of various objects. A \ac{DT} is a real-time evolving digital duplicate of a physical object or a process that contains all of its history \cite{he2018surveillance}. Its implementation involves massive real-time multi-source data collection, analysis, inference, and visualisation. 
Although the \ac{DT} technology already exists in some industrial applications supported by \ac{5G} or even 4G \cite{Tao18Survey}, it has not been widely adopted in other sectors, and has not reached its full potential. The need for high throughput (100 Gbs), reliable (99.999\%) and pervasive communication is one of the bottlenecks in realising \ac{DT}'s potential, requiring beyond-\ac{5G} technologies. 
Therefore, \ac{6G} can be considered as an enabler for massive adoption of \acp{DT}. 
%

%
The popularity of \ac{DT} depends on the popularity and necessity of its applications. Potential high-connectivity-demanding and rapidly emerging applications of \ac{DT} ranging from aerospace, which has very high mobility, to Industry 4.0, which has very high number of devices in a location, and healthcare with high reliability requirement, could be a major driver toward the development of \ac{6G} \cite{Tao18Survey}. In this paper we also argue that the network itself can have its \ac{DT} which will be an important application of \ac{DT}. In addition, as will be discussed in the next sections, the \ac{DT} technology itself integrated with \ac{AI} could act as a facilitator toward this development.

In this paper, we aim to highlight and further explore this relationship between \ac{6G} and \ac{DT}.
Section II will further describe \ac{DT} and its features and requirements. Potential application of \ac{DT} in future communication networks and in particular \ac{6G} are presented in Section III. \ac{6G} as a facilitator for wide adoption of \acp{DT} is then discussed in Section IV. Finally, a road-map for future research directions and some concluding remarks are presented in Section V and VI, respectively. 

\section{Digital Twin}

The term ``\acl{DT}" was first coined  by Grieves in 2003 \cite{grieves2014digital}. The technology became more popular after the emergence of Industry $4.0$ (in 2016) as it enabled integration of digital manufacturing and cyber-physical systems. 

A \ac{DT} can be defined as a ``virtual representation of an asset, providing both a historical ledger of the asset’s previous states, and real-time data on the asset’s current state". The asset can be an object, a machine, a process, or even a system. 
A \ac{DT} requires a real-time bidirectional connection with its \ac{PT}. It should be clarified that \ac{DT} is more than an avatar, a surveillance, a simulation, or a simple model. An avatar is a limited replica of the physical asset without any possibility of controlling the asset. In addition, the bidirectional connection with the \ac{PT}, makes a \ac{DT} more sophisticated and capable than a surveillance system. Unlike \textit{simulation}, a \ac{DT} ideally represents an actual asset with as little assumptions or simplifications as possible (except those that are required to digitally encode the physical asset involved).

While a \ac{DT} can benefit from \ac{AR}/\ac{VR}/\ac{MR} for visualisation purposes, it is different from \emph{augmented virtuality}. The main focus and goal of \emph{augmented virtuality} is representation and human interaction. However, \acp{DT} mainly focus on maintaining the full history and up-to-date information of the assets/systems to facilitate intelligent and data-supported decision making \cite{Maier_AV20}. In the following, we briefly discuss key features and specifications of \ac{DT} as well as relevant standardisation activities and challenges.

\subsection{Key Features of \acl{DT}}
\begin{figure}
    \centering
    \includegraphics[width=0.85\columnwidth]{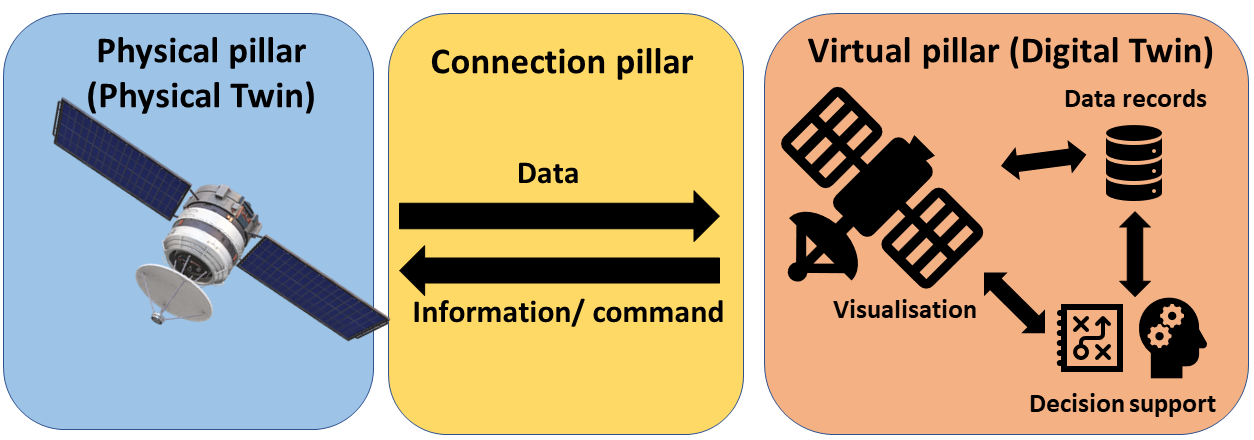}
    \caption{Pillars of a \acl{DT} System}
    \label{fig:dt-def}
\end{figure}

\subsubsection{Pillars}
A \ac{DT} system is composed of three pillars i.e., physical, digital/virtual, and connection pillars \cite{Tao18Survey}. Figure \ref{fig:dt-def} presents an example of a \ac{DT} system and its pillars. The physical pillar, representing the \ac{PT}, is the actual asset which is the basis of the digital model and the source of its data.  The virtual/digital pillar, or equivalently the \ac{DT}, is the host of the data models, historical data of the \ac{PT}, decision support, \ac{AI}, and visualisations. The \ac{DT} is capable of sending control commands to the physical pillar. The connection pillar between a \ac{PT} and a \ac{DT} is the communication bridge that allows for the exchange of data and control commands among them. 
The connection pillar is not necessarily symmetric as the flow of data in each direction, \ac{PT}-to-\ac{DT} versus \ac{DT}-to-\ac{PT}, requires different levels of \ac{QoS}.
In this paper, the phrase \textit{\ac{DT} system} refers to a complete system consisting of all three pillars, while the term \textit{\ac{DT}} only indicates the digital pillar of the system. It should be emphasized that the \ac{DT} or digital pillar of any physical asset is only meaningful when it is functioning as part of a complete \ac{DT} system. 

\subsubsection{Modularity} 
Modularity is the key to interoperability and interchangeability. Modularity enables the system to evolve as the technology on each component evolves. In a modular system the interfaces are standardised and therefore the components can be replaced, due to technology upgrade or maintenance seamlessly.

A \ac{DT} can be highly modular \cite{rasheed2020digital}. It is possible to create a \ac{DT} for every single component of an asset and create a mega-\ac{DT} by interconnecting the smaller \acp{DT} representing those components. This feature enables rapid reproduction of processes and knowledge transfer. 
Modularity of a \ac{DT} allows creating hybrid simulation and prototyping systems. In such systems, the \acp{DT} of existing physical subsystems are combined with a simulation of subsystems which still do not have a corresponding \ac{PT}. A hybrid system can accelerate the design, development, and prototyping of new products and services. It can also enable performance testing of the physical subsystems in a virtual replica of the target application environment (within the boundaries of the data model used to represent the related \acp{PT}).
\subsubsection{Remote Intelligence}
The capability to apply remote intelligence to enhance the operation of the \ac{PT} is another important feature of a  \ac{DT}.
A resource-limited physical device or an old machinery can become more efficient or intelligent by running data analysis, AI algorithms, or even conventional optimisation and/or analysis algorithms on its \ac{DT} which can be located at the edge, or in the cloud.
\subsection{Standardisation}
Modularity feature of \ac{DT} enables creation of mega-\acp{DT} by rapid reproduction of processes from \acp{DT} of different components. This necessitates interoperability among these components and therefore highlights the importance of \ac{DT} standardisation. The current activities on \ac{DT} standardisation are focusing on data collection, storage, and exchange \cite{jacoby2020digital}. Microsoft\footnote{Commercial products and companies mentioned in this paper are merely intended to foster understanding. Their identification does not imply recommendation or endorsement by the National Institute of the Standards and Technology.} is developing a programming language independent data management model based on JavaScript Object Notation for Linked Data (JSON-LD) called \emph{Digital Twin Definition Language} (DTDL). DTDL is used for data management of \acp{DT} that are deployed using Microsoft Azure. Although DTDL addresses the interoperability challenges on Azure-based \acp{DT}, lack of comprehensive standardisation could affect \ac{DT} adoption especially for their deployment on the edge \cite{rasheed2020digital}. 

Another candidate for widespread standardisation of \acp{DT} could be the functional mockup interface (FMI) (https://fmi-standard.org/). It is currently a free standard that enables building \acp{DT} of different \acp{PT} using combinations of XML and C codes.

Several other relevant existing standards for example, Object Linking and Embedding for Process Control (OPC) Unified Architecture (OPC-UA), which is a standard for machine-to-machine communication can be leveraged towards \ac{DT} standardisation. OPC-UA can be used to connect the components of the \ac{PT} and then the communication links between the \ac{PT} and the \ac{DT} can utilise existing application programming interfaces (APIs) like the REpresentational State Transfer (REST) API. All these standards along with newly defined ones can be brought together to define a unified set of standards for \acp{DT}.

\section{Digital Twin of Communication Networks}

So far the \ac{DT} technology has been adopted in manufacturing, healthcare, and aviation 
\cite{rasheed2020digital}. 
In telecommunications industry, companies like Spirent Communications and British Telecommunication (BT) have started developing \acp{DT} for \ac{5G} network components. These activities will pave the way for full adoption of \ac{DT} in \ac{6G}.

Similar to its application in other industries, using \ac{DT} of a telecommunication network or any of its components can significantly improve network design and maintenance. 
This directly affects network's life cycle as discussed in the remainder of this section.

\subsection{Network and \ac{DT}'s life cycle}
\begin{figure*}
    \centering
    \includegraphics[width=.75\textwidth]{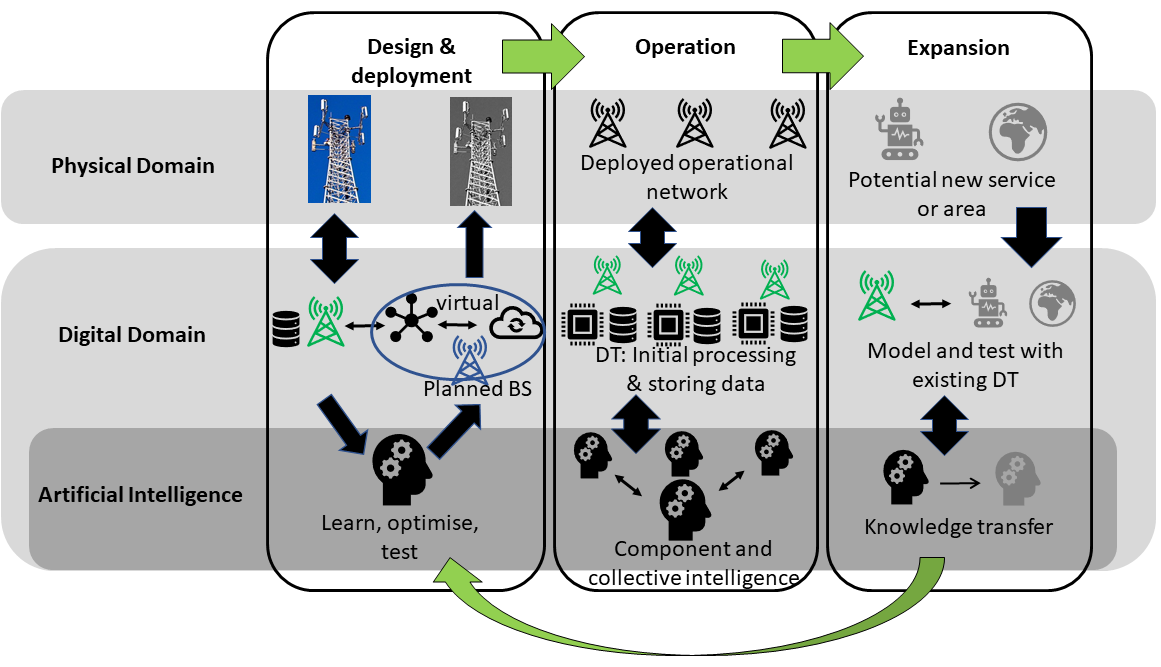}
    \caption{A network's life cycle using \ac{DT}. The grey icons like base station indicate that they have not been deployed yet, and the action results of the other side of arrow leads to their existence.}
    \label{fig:digandphy}
\end{figure*}

The evolving digital replica of a network that is provided by its \ac{DT} can assist in the design, deployment, operation, and expansion phases of a network
. This is shown in Figure \ref{fig:digandphy} and further illustrated in the following.

\subsubsection{Design and deployment} 
In the era of \acp{DT}, simulation and model-based network design is replaced by an analytics-supported design process. Modularity of \acp{DT} enables network designers to exploit the existing knowledge on \acp{DT} of various networks' components. Engineers will then be able to design the communication network by creating a hybrid-simulation environment using the modularity feature. 
As observed in Figure \ref{fig:digandphy}, the design and deployment phase starts with a physical component of the network such as a \ac{BS} (shown in blue highlight) and its \ac{DT}. The rest of the network is designed in digital domain using \ac{AI}.
Once the design process, test, and verification is complete through analytics in the hybrid system, the deployment phase starts (the \ac{BS} shown in grey highlight). As different sections (or subsystems) of the network are deployed, their \acp{DT} will be created and merged with the hybrid simulation environment. By the end of the deployment phase, the hybrid system becomes a complete \ac{DT}. The key difference in this methodology compared to existing network simulation and planning tools used in \ac{5G} and earlier generations, including general ones and proprietary tools, is that, \ac{DT}-based systems are connected to deployed physical subsystems and the whole system evolves as the deployment proceeds. 

\subsubsection{Smart operation, maintenance, and resilience}
Phase two deals with the operation of the network as shown in Figure \ref{fig:digandphy}'s operation phase. 
Here, an \ac{AI}-enabled \ac{DT} optimises the operational parameters of the network based on the real-time data and the knowledge generated through prior experience. Resilience is the ability of the network to maintain an acceptable level of service in the event of various faults and challenges to normal operation \cite{Hamed_Res18}. 
\textit{Resilience} cannot be achieved if the network is not prepared for potential disruptions. \ac{AI} can check all possible \textit{what-if} scenarios and choose the network configuration which guarantees operation with the highest \ac{QoS}. This is a step beyond what is known as \ac{SON}. To achieve real resilience, the \ac{AI} in the \ac{DT} acts beyond self-optimisation and self-healing, and performs prediction and strategic planning. In \ac{SON}, questions like placement of the required intelligence and the coordination with legacy systems still remains unclear. \ac{DT} modularity supports intelligence at the edge, federated learning, and transfer learning to provide maximum resilience \cite{park2019wireless}. Basically modularity will bring the flexibility to add and remove crucial components at different locations and essentially provide the redundancy as and when needed. It is true that redundancy improves the resilience, but it also increases cost and overhead. Our point is that with \ac{DT} modularity, intelligence is supported and intelligence predicts potential disruptions. Predicted disruptions can be taken care of before happening and the system will be resilient without the need of having costly redundant copies for each and every component. Also,  additional sensors and edge computation can be used to create \acp{DT} for legacy equipment. 

Maintenance, prediction, and strategic planning can be better clarified with the following toy example. The equipment used in a network have a certain lifetime beyond which they either need maintenance or replacement. The estimated lifetime is normally provided by the manufacturer. However, in practice, the actual lifetime could differ from this estimate based on the working load and physical condition of the operating location. The \ac{AI} on the network's \ac{DT} or each of its component's \ac{DT} are capable of learning each component's lifetime from the manufacturer data, the real-time data received from the \ac{PT}, as well as other external factors. As a result, the \ac{DT} can modify the network's working conditions to maximise the lifetime of different equipment and/or efficiently schedule maintenance time. In real scenarios, other than this toy example, the optimisation should consider optimal service and other important criteria too.
Using \ac{AI} facilitated by \ac{DT} to support network's operation, enables the network to predict its disruptions caused by components failure or other sources, proactively respond to them, and/or prevent them before happening. 
 \begin{figure*}
    \centering
    \includegraphics[width=6.5 in]{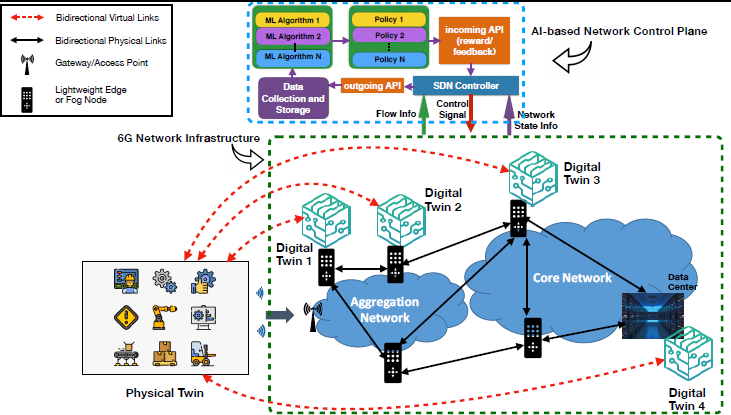}
    \caption{Communication of \ac{PT} and \acl{DT} over a 6G network}
    \label{fig:dtover6G}
\end{figure*}
\subsubsection{Knowledge transfer, and robust expansion}
The last phase of most products in manufacturing is dismissal phase and release of a new product based on the changes in the market and the lessons learnt from the existing product. In telecommunications domain we can translate it to network expansion to new domains, geographical locations, and/or providing new services; for example, using \ac{DT} of a \ac{5G} network to transfer knowledge for the design phase of \ac{6G}. Disconnected twins of components' or the complete network can be used for the design of new networks and testing new services. Additionally, operators can monetise their experience by selling the data and the created knowledge via disconnected twins \cite{altun2019liberalization}. As shown in Figure \ref{fig:digandphy}, this phase closes the network life cycle loop. 

\subsection{DT of the next-generation of networks}
As \ac{5G} has already reached its deployment phase and its standardisation has been almost completed, \ac{DT}-based design and operation of networks can show its benefit mostly in \ac{6G}. Using \ac{DT}-based approach, \ac{6G} can be designed and standardised in a more data-oriented fashion. In the operation phase, \ac{6G} will be capable to handle its own \ac{DT} while the massive overhead created by the \ac{DT} of the network cannot be handled by \ac{5G} while supporting its high throughput and/or ultra delay sensitive usual services. \ac{6G}'s high \acp{KPI} in addition to its synergy with \ac{AI}, will enable it to support the additional overhead to have its own \ac{DT}. In the next Section, we present how \ac{6G} can support other \acp{DT}.

\section{6G as an Enabler of Digital Twin}

As discussed so far, a \ac{DT} is implemented using a combination of a simulation of the physical system and a means to \emph{communicate} all the data generated by the physical system to its \ac{DT} and, the \ac{AI}-processed, command and control from \ac{DT} to the physical system. The \emph{communication} part involved in the successful synergy between a \ac{DT} and its corresponding \ac{PT} has to support ultra-reliable, real-time (or semi-real-time depending on the application), and high \ac{QoS} communication.

At present, \ac{DT} technology is mainly used in industrial plants and it is supported by \ac{5G} or earlier generations of communication protocols. It is quite conceivable that wide adoption of this technology results in higher capacity demands as well as new scenarios beyond the capabilities of \ac{5G}. Next we discuss some of these scenarios.

General Electric is one of the pioneers in using \ac{DT} technology in manufacturing. According to the company \$1.6B has been saved by early detection of industrial components failure through continuous remote monitoring of assets \cite{GE}. In such scenarios, network reliability is extremely important, and full wired connection is not an option due to its complexity of installation and high cost. \ac{6G} promising a reliability of 99.999\% translates to a yearly downtime of 5 minutes as compared to the 8-and-half hours of downtime with \ac{5G}'s 99.9\% reliability. Therefore, for future massive-scale industrial IoT applications facilitated by \acp{DT}, a \ac{6G} network is more advantageous than its \ac{5G} counterpart.

Figure \ref{fig:dtover6G} gives a schematic detail of how a \ac{PT} in an industrial \ac{IoT} use-case can have different \acp{DT} for each of its components distributed over the cloud and the edge, supported by a 6G network infrastructure. The \ac{PT}, a factory with different physical systems, is modelled as a combination of several \acp{DT}. The \acp{DT} are distributed in various cloud and edge servers. The red dotted-lines represent logical bidirectional connections between the \ac{PT} and the \acp{DT}. 
The network infrastructure as depicted in Figure \ref{fig:dtover6G} has a fully automated control plane. This control plane can orchestrate the network using \ac{AI} algorithms that are continuously trained by the network data. 
\ac{AI}-supported autonomous operation of this complex system (mega-\ac{DT}) requires near-perfect connection between the \acp{DT} on the edge and the cloud servers. \ac{6G} can support this mega-\ac{DT} with millisecond latency, 100 GB/s data rate and 99.999\% reliability.

A \ac{DT} system can benefit from integrated
modern visualisation technology in order to display complex data types to the users. To enable that, many networked data-collection devices e.g., high-resolution cameras, are required and this has to be enabled in the edge networks \cite{he2018surveillance}. Processing ultra-high-definition videos along with complex AI algorithms like Deep Learning would require significant processing power localised in a single or few nodes. A more feasible solution is to enable federated \ac{AI} where different components of the \ac{AI} algorithms can be distributed over the network nodes \cite{park2019wireless}. For example, a deep neural network can have some of its inputs/outputs in the low-complexity edge nodes while hidden neurons reside in the cloud with more processing power. These spatially distributed components of the neural network require ultra-reliable communication to avoid erroneous training and output. Although federated \ac{AI} has been implemented using \ac{5G} in small scale, its larger and more complex deployment could require 99.999\% reliability and one millisecond latency of \ac{6G} \cite{park2019wireless}.

Furthermore, due to the modularity feature of  \acp{DT},
they may not be localised in either a single node or a small subset of nodes \cite{rasheed2020digital}. As a result, the data associated with the \acp{DT} and the \ac{AI} that operates on these data may have to be distributed over the cloud and/or several edge servers across the network. 
Seamless communication among these distributed \acp{DT}, computation associated with the distributed \ac{AI} operating on these \acp{DT}, and maintaining security and integrity of these data is a challenge. One solution is using Blockchain-based transactions between these nodes. However, high transaction throughput requirement (i.e., 10,000 transactions per second and millisecond latency) of private and consortium Blockchains can only be satisfied by \ac{6G}-level of \ac{QoS} \cite{viswanathan2020communications}.

\section{Future Research Directions}

Having introduced the concept of \ac{DT} in telecommunications and its potential roles in setting up and transforming \ac{6G} networks, both as an enabler and a use-case, in this section we enumerate several key research directions related to this combined field.

\subsection{DT Ownership Issues}
\ac{DT} ownership is a challenging issue with technical, financial, and legal aspects. The challenge is mainly caused by the potential difference in the ownership of the physical entity and the \ac{DT} platform. A simplified example 
of this scenario is the common fitness trackers. A fitness tracker device is owned by an individual, while the generated data is owned by and stored on the application provider's cloud. Normally, the individual can only access the data via a specific application interface without the option of exporting the data. However, the individual can disconnect the fitness tracker or simply stop using it at any time. Since a \ac{DT} contains more detailed information and needs to be always connected with the physical object, ownership issues must be clarified. This is especially important considering the General Data Protection Rules (GDPR) introduced in the European Union. In \cite{altun2019liberalization}, the authors considered home appliances in an \ac{IoT} scenario. The owner of an appliance, if also interested in full ownership of the data, can buy, install and maintain its \ac{DT} on his/her home gateway. While this is a viable option, it requires owning a gateway with sufficient storage capacity and security. Alternatively, the appliance owner can rent cloud/fog/edge services to install and maintain the \ac{DT}. Therefore, the ownership issue will go hand-in-hand with cloud/fog/edge computing and placement challenges. 
In \cite{altun2019liberalization} the \ac{IoT} devices are connected to the home network and the ownership scenario will be more complicated in industrial settings with the use of \ac{6G}.
The investigation of more complicated ownership-related scenarios, especially for process or system \acp{DT} with multiple components owned by different entities, remains open for further research. 

\begin{figure*}
    \centering
    \includegraphics[width=.8\textwidth]{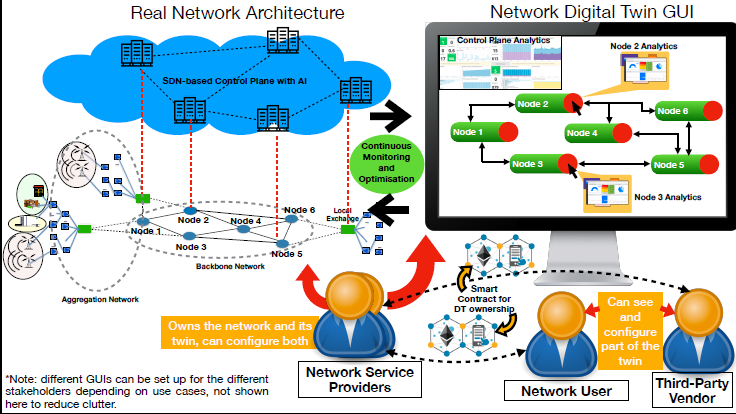}
    \caption{Digital Twin for an entire network}
    \label{fig:netDT}
\end{figure*}

\subsection{Ultra-Low-Latency and Reliable Communication between DT and PT}
As mentioned previously, a seamless real-time data exchange between the \ac{DT} and the \ac{PT} is a necessary condition to define a \ac{DT} system. Significant amount of data has to be continuously and reliably exchanged between the pair. The software tools, data analytics modules, and the data that makes the \ac{DT} an appropriate clone of the \ac{PT} should mostly be stored in the cloud. However, for some critical use-cases e.g., the \ac{DT} of a remote-surgery system, implementation in the edge might be preferred \cite{Laaki2019}. 

Whatever the scenario, it is anticipated that most \ac{DT} implementations would require Ultra-Low-Latency and Reliable Communication between the \ac{DT} and its \ac{PT}. Recent research studies have established the importance of Ultra-Low-Latency and Reliable Communication for some future applications, and reported the development of technologies and algorithms that could make that achievable \cite{viswanathan2020communications, saad2019vision}. However, further breakthroughs across all protocol layers of the network are still needed to achieve strict latency and reliability requirements.
%
\subsection{Federated DT in the Cloud/Edge}
Resources such as power, storage, high-speed memory are sometimes constrained in today's networks. Therefore, significant resource management is necessary to sustain a technology like the \ac{DT}, which includes communication, data analysis, and AI-based computation. To accommodate various use-cases of \ac{6G}-and-beyond-networks, it is anticipated that a large percentage of the computation (including AI algorithms) and storage is moved to the edge of the network \cite{park2019wireless}. The trend will be similar if \ac{6G}-and-beyond-networks have to support massive adoption of \ac{DT} technology. Having said that, it will be almost imperative that several backend solutions enabling a \ac{DT} for a particular \ac{PT} need to be hosted in multiple data centers in the cloud and/or edge. 

There are several reasons for the need to do this distribution or even replication of \acp{DT}. First of all, the storage and computing facilities of the servers in the cloud or the edge may pose system-level limitations to host a \ac{DT} in one place. This might create unnecessary performance bottlenecks. Secondly, there might be failures in the servers or network links which might hamper the seamless connectivity between a \ac{PT} and its \ac{DT}. Therefore, it is pragmatic to distribute multiple copies of \acp{DT} all over the cloud and/or the edge servers as illustrated in Fig. \ref{fig:dtover6G}. 

Several components of the cloud and/or edge distributed \acp{DT} need to communicate with one another to exchange data and/or train \ac{AI} models to establish automated and intelligent operations. This can be termed as \emph{federated DT} similar to the concept of federated learning as proposed in \cite{park2019wireless}. It is a challenging task to run such forms of synchronised and collaborative AI algorithms over the nodes of the network. This is still an open research area. 

\subsection{DT of an Entire Network}
As mentioned before, the \ac{DT} technology has not been utilised much for telecommunication networks. Today's telecommunication networks are getting softwarised, owing to new trends like \ac{SDN} and \ac{NFV}. 
The advent of \ac{AI} in addition to the network softwarisation is further pushing the drive towards automated and autonomous telecommunication networks. Therefore, apart from the physical infrastructure (i.e., transceivers, antennas, optical fibers, filters, etc.) most of the other network components can be implemented as cloud-native software. 

This would constitute a paradigm shift in terms of how the future networks can be managed and used, if a composite \ac{DT} of an entire network can be created. If the \acp{DT} of the physical components of the networks (i.e., transceivers, switches, links) can be implemented, they can be nicely intertwined along with the other softwarised components of the network to form a composite \ac{DT} of the network. Just like a massive manufacturing unit or a giant space shuttle can be troubleshot and managed by tuning several parameters on their \acp{DT}, an entire network can also be managed, upgraded, and troubleshot using its \ac{DT}. Several network services and new technologies pertaining to the networks can also be tested and pre-implemented on these massive-scale network \acp{DT} before deploying in the real networks. 

Figure \ref{fig:netDT} captures our vision towards enabling the \ac{DT} of an entire network. It also highlights some of the related research issues like network monitoring and troubleshooting using AI-based analytics and ownership issues using \emph{smart contracts} hosted in a Blockchain. 

\subsection{Experimental Investigation of \acp{DT}}
The development of a complete LTE network using commercially available software components such as Amarisoft\textsuperscript{TM} LTE 100 eNodeB, UE from software radio systems (srsUE\textsuperscript{TM}) and a generic RF front end has been document in \cite{qomex}. This network was entirely switched ON/OFF using a python and Linux based code. The code would turn on the LTE network, stream a YouTube video, collect data from the video for analysis in real-time and plot various performance curves. A similar type of setup can prove to be a suitable starting point for an experimental investigation of the \ac{DT} of a network. More developments would still be required to build a Graphical User Interface (GUI) to visualise the operations of all components, and to set up real time connections between the graphical representations of the \acp{DT} and the \acp{PT}.

\section{Conclusions}
In this paper we discussed the application of \ac{DT} in networking and presented its potential relationship with \ac{6G}. While \ac{6G} can facilitate realisation and adoption of \ac{DT} in several industries by providing the required levels of reliability and speed, \ac{DT} integrated with \ac{AI} can also facilitate \ac{6G} networks design, deployment, and operation. This approach can have significant impact on achieving high network resilience. Additionally, demanding applications of \ac{DT} ranging from aerospace to Industry 4.0 and healthcare, could be a major driver towards the development of 6G. Potential \ac{DT}-related research directions have also been highlighted in the paper. 

\begin{acronym} 
\acro{5G}{the fifth generation of mobile networks}
\acro{6G}{sixth generation of mobile networks}
\acro{ACO}{Ant Colony Optimization}
\acro{AI}{Artificial Intelligence}
\acro{AR}{Augmented Reality}
\acro{BB}{Base Band}
\acro{BBU}{Base Band Unit}
\acro{BER}{Bit Error Rate}
\acro{BS}{Base Station}
\acro{BW}{bandwidth}
\acro{C-RAN}{Cloud Radio Access Networks}
\acro{CAPEX}{Capital Expenditure}
\acro{CoMP}{Coordinated Multipoint}
\acro{CR}{Cognitive Radio}
\acro{D2D}{Device-to-Device}
\acro{DA}{Digital Avatar}
\acro{DAC}{Digital-to-Analog Converter}
\acro{DAS}{Distributed Antenna Systems}
\acro{DBA}{Dynamic Bandwidth Allocation}
\acro{DC}{Duty Cycle}
\acro{DL}{Deep Learning}
\acro{DSA}{Dynamic Spectrum Access}
\acro{DT}{Digital Twin}
\acro{FBMC}{Filterbank Multicarrier}
\acro{FEC}{Forward Error Correction}
\acro{FFR}{Fractional Frequency Reuse}
\acro{FSO}{Free Space Optics}
\acro{GA}{Genetic Algorithms}
\acro{HAP}{High Altitude Platform}
\acro{HL}{Higher Layer}
\acro{HARQ}{Hybrid-Automatic Repeat Request}
\acro{IoT}{Internet of Things}
\acro{KPI}{Key Performance Indicator}
\acro{LAN}{Local Area Network}
\acro{LAP}{Low Altitude Platform}
\acro{LL}{Lower Layer}
\acro{LOS}{Line of Sight}
\acro{LTE}{Long Term Evolution}
\acro{LTE-A}{Long Term Evolution Advanced}
\acro{MAC}{Medium Access Control}
\acro{MAP}{Medium Altitude Platform}
\acro{MIMO}{Multiple Input Multiple Output}
\acro{ML}{Machine Learning}
\acro{MME}{Mobility Management Entity}
\acro{mmWave}{millimeter Wave}
\acro{MNO}{Mobile Network Operator}
\acro{MR}{Mixed Reality}
\acro{NASA}{National Aeronautics and Space Administration}
\acro{NFP}{Network Flying Platform}
\acro{NFPs}{Network Flying Platforms}
\acro{NTN}{Non-terrestrial networks}
\acro{NFV}{Network Function Virtualisation}
\acro{OFDM}{Orthogonal Frequency Division Multiplexing}
\acro{OSA}{Opportunistic Spectrum Access}
\acro{PAM}{Pulse Amplitude Modulation}
\acro{PAPR}{Peak-to-Average Power Ratio}
\acro{PGW}{Packet Gateway}
\acro{PHY}{physical layer}
\acro{PSO}{Particle Swarm Optimization}
\acro{PT}{Physical Twin}
\acro{PU}{Primary User}
\acro{QAM}{Quadrature Amplitude Modulation}
\acro{QoE}{Quality of Experience}
\acro{QoS}{Quality of Service}
\acro{QPSK}{Quadrature Phase Shift Keying}
\acro{RF}{Radio Frequency}
\acro{RN}{Remote Node}
\acro{RRH}{Remote Radio Head}
\acro{RRC}{Radio Resource Control}
\acro{RRU}{Remote Radio Unit}
\acro{SU}{Secondary User}
\acro{SCBS}{Small Cell Base Station}
\acro{SDN}{Software Defined Network}
\acro{SNR}{Signal-to-Noise Ratio}
\acro{SON}{Self-organising Network}
\acro{TDD}{Time Division Duplex}
\acro{TD-LTE}{Time Division LTE}
\acro{TDM}{Time Division Multiplexing}
\acro{TDMA}{Time Division Multiple Access}
\acro{UE}{User Equipment}
\acro{UAV}{Unmanned Aerial Vehicle}
\acro{USRP}{Universal Software Radio Platform}
\acro{VR}{Virtual Reality}
\acro{XAI}{Explainable Artificial Intelligence}
\end{acronym}
\balance
\bibliographystyle{IEEEtran}
\bibliography{MyRef}

\section*{Biographies}
\vspace{-0.3in}
\begin{IEEEbiography}
{Hamed Ahmadi (SM'15)}
is an assistant professor in the department of Electronic Engineering at University of York, UK. He received his Ph.D. from National University of Singapore in 2012. Since then he worked at different academic and industrial positions in Ireland and UK. His research interests include design, analysis, and optimization of wireless communications networks, application of machine learning and Blockchain in wireless networks. 
\end{IEEEbiography}
\vspace{-0.3in}
\begin{IEEEbiography}
{Avishek Nag (SM'18)}
 is currently an Assistant Professor in the School of Electrical and Electronic Engineering at University College Dublin (UCD) in Ireland. Dr Nag received his PhD degree from the University of California, Davis. He worked as a research associate at the CONNECT Centre for future networks and communication in Trinity College Dublin, before joining UCD. His research interests include the application of optimisation theory, AI, and Blockchain in telecom networks. Dr Nag is the outreach lead for Ireland for the IEEE UK $\&$ Ireland Blockchain Group. 
\end{IEEEbiography}
\vspace{-0.3in}
\begin{IEEEbiography}
{Zaheer Khan}
received Dr.Sc. degree in electrical engineering from the University of
Oulu, Finland, in 2011. He was a Research Fellow/Principal Investigator with the University of Oulu, from 2011 to 2016, where he is currently an Adjunct Professor. His research interests include the implementation of advanced signal processing and wireless communications algorithms on Xilinx FPGAs and Zynq system-on-chip (SoC) boards, application of game theory to model distributed wireless networks, and wireless signal
design. He was a recipient of the Marie Curie Fellowship, from 2007 to 2008.
\end{IEEEbiography}
\vspace{-0.3in}
\begin{IEEEbiography}
{Kamran Sayrafian (SM'06)}
is a Senior Scientist and Program Lead at the Information Technology Laboratory of the National Institute of Standards and Technology (NIST) located in Gaithersburg, Maryland. He is leading several strategic projects related to the application of IoT in Healthcare. Dr. Sayrafian is also an affiliate Associate Prof. of Concordia
University in Montreal, Canada since 2016. He is currently member of the Editorial Board of the IEEE Wireless Communication Magazine. His research interests include IoT-Health, micro energy-harvesting technology, mobile sensor networking and RF-based positioning. He has published over 120 conference and journal papers, and book chapters in these areas, and has been the recipient of the IEEE PIMRC 2009, SENSORCOMM 2011, IEEE CSCN 2018 and IEEE EuCNC 2019 best paper awards. Dr. Sayrafian is the co-inventor/inventor of four U.S. patents.
\end{IEEEbiography}
\vspace{-0.3in}
\begin{IEEEbiography}
{Susanto Rahardja (F'11)}
 is currently with Institute for Infocomm Research. His research interests are in multimedia, signal processing, wireless communications, discrete transforms, machine learning and signal processing algorithms and implementation. Dr Rahardja has more than 15 years of experience in leading research team for media related research that cover areas in Signal Processing, Media Analysis, Media Security and Sensor Networks. He has published more than 300 papers and has been granted more than 70 patents worldwide out of which 15 are US patents. Dr Rahardja was past Associate Editors of IEEE Transactions on Audio, Speech and Language Processing and IEEE Transactions on Multimedia, past Senior Editor of the IEEE Journal of Selected Topics in Signal Processing, and is currently serving as Associate Editors for the Elsevier Journal of Visual Communication and Image Representation and IEEE Transactions on Multimedia.\end{IEEEbiography}

\end{document}